\documentclass[twocolumn]{revtex4}
\usepackage{graphicx,amssymb}
\usepackage{dcolumn}
\usepackage{bm}
\usepackage{longtable}
\usepackage{epsfig}

\def\prl{Phys. Rev. Lett. }
\def\prb{Phys. Rev. B }
\def\apl{Appl. Phys. Lett. }
\def\jap{J. Appl. Phys. }
\def\jjap{Jpn. J. Appl. Phys.  }

\def\nimb{Nucl. Instrum. Methods Phys. Res. B }
\def\tsf{Thin Solid Films }
\def\ss{Surf. Sci. }
\def\aps{Appl. Surf. Sci.  }

\begin{document}

\title{Study of low energy Si$_5^-$ and Cs$^-$ implantation 
induced amorphisation effects in Si(100) } 

\author {H. P. Lenka,$^1$ B. Joseph,$^{1,}$\footnote{Present address: Istituto 
Tecnologie Avanzate, Contrada Milo 91100, Trapani, Italy} 
P. K. Kuiri,$^1$ G. Sahu,$^1$ P. Mishra,$^2$ D. Ghose,$^2$ and 
D. P. Mahapatra$^{1,}$\footnote {Author to whom any correspondence should be 
addressed; electronic mail: dpm@iopb.res.in}} 
\affiliation {$^1$ Institute of Physics, Sachivalaya Marg, Bhubaneswar - 
751005, India \\ $^2$ Saha Institute of Nuclear Physics, 1/AF Bidhan Nagar, 
Kolkata - 700064}

\begin{abstract}
The damage growth and surface modifications in Si(100),
induced by 25 keV Si$_5^-$ cluster ions, as a function 
of fluence, $\phi$, has been studied using atomic force 
microscopy (AFM) and channeling Rutherford backscattering 
spectrometry (CRBS). CRBS results indicate a nonlinear 
growth in damage from which it has been possible to get 
a threshold fluence, $\phi_0$, for amorphisation as 
$2.5\times 10^{13}$ ions cm$^{-2}$. For $\phi$ below 
$\phi_0$, a growth in damage as well as surface roughness 
has been observed. At a $\phi$ of $1\times 10^{14}$ 
ions cm$^{-2}$, damage saturation coupled with a much 
reduced surface roughness has been found. In this case 
a power spectrum analysis of AFM data showed a significant 
drop, in spectral density, as compared to the same 
obtained for a fluence, $\phi < \phi_0$. This drop, 
together with damage saturation, can be correlated with 
a transition to a stress relaxed amorphous phase. 
Irradiation with similar mass Cs$^-$ ions, at the same 
energy and fluence, has been found to result in a 
reduced accumulation of defects in the near surface region
leading to reduced surface features.
\end{abstract}
\date{\today}
\maketitle

\section{Introduction}

Cluster ion implantation can be regarded as a forerunner 
technology as compared to the conventional ion implantation 
technique used to dope sub-micron devices \cite{Popok1,Yamada1}. 
Using cluster ions very shallow implantation can be achieved 
at very low energy. However, with cluster implantation, 
nonlinear effects arising in the energy loss processes, as 
a result of the correlated motion of the constituent atoms, 
play an important role in deciding the defect structure near 
the target surface. In addition to resulting in a nonlinear 
growth in subsurface damage, cluster ion impact, through 
sputtering, can also results in kinetic roughening and 
smoothening of the surface exposed \cite{ichimura}. 
In view of all this, there has been a lot of activities 
involving low energy cluster ion irradiation related to 
nonlinear sputtering \cite{anderson-Samartsev}, nonlinear 
damage and defect production \cite{Chu-Liu,Cn-nimb07,titov}, 
along with the formation of various kind of surface features 
\cite{prasa,nordlund1,Choi,Yamada2,Popok3}. 

In connection with the above, Si, presents itself as a
very important material where low energy cluster ions 
can be used for shallow implantation, of interest to 
technology. In some earlier work, contrary to common 
expectation, amorphisation upon ion irradiation has 
been shown to start from the surface rather than the ion
projected range \cite{nakajima}. Results of Molecular dynamics (MD)
simulations with 5 keV Si, show that the ion impacts produce unrelaxed  
amorphous patches that have a fast quenched, liquid like 
structure \cite{delaRubia}. 
With increase in ion fluence these regions overlap 
producing a continuous amorphous layer \cite{rimni}. In 
fact, with increase in ion fluence, there is a superlinear 
growth of amorphous volume fraction with a lot of stress 
build up in the matrix. 
At high fluence there is an abrupt transition to a state 
with a flat amorphous-to-crystalline (a/c) interface 
\cite{holland,bai}. In such a case, out of plane plastic 
flow with a reduction in the in-plane stress have been 
observed \cite{volkert}. All this suggest that ion
irradiation induced amorphisation in Si is more like a 
{\it phase transition}, initiated by a spontaneous collapse 
of the damaged region. 
Very recent MD simulations carried out by Marqu\'{e}s 
{\it et al} show it to be initiated by a high concentration 
of {\it interstitial-vacancy (IV) pairs} or {\it bond defects}, 
formed in the system \cite{pelaz}. Similar results have 
also been shown by Nord {\it et al} \cite{nordlund2} who have 
pointed out that the subsequent transition resulting in a 
uniform amorphous layer is neither a complete homogeneous 
nor a complete heterogeneous mechanism.
This makes Si an ideal system to study using low energy 
cluster ions where such a transition to a complete 
amorphous state is expected at a lower fluence, primarily 
because of overlapping of collision cascades coming from 
constituent atoms. 

In the present paper we show some results of a systematic 
study of the subsurface damage produced and the surface 
features generated in Si(100), from Si$_5^-$ and a similar 
mass Cs$^-$ ion implantation at 25 keV. 
Channeling Rutherford backscattering spectrometry (CRBS) 
and 
Atomic force microscopy (AFM) have been used for sample
characterization. Increase in cluster ion fluence has been
found to result in a nonlinear growth and saturation in 
damage leading to amorphisation. The transition to an 
amorphised state is found to be associated with a significant
drop in the power spectral density of AFM data which 
initially increases with increase in fluence.

\section{Experiment}

\begin{figure*}
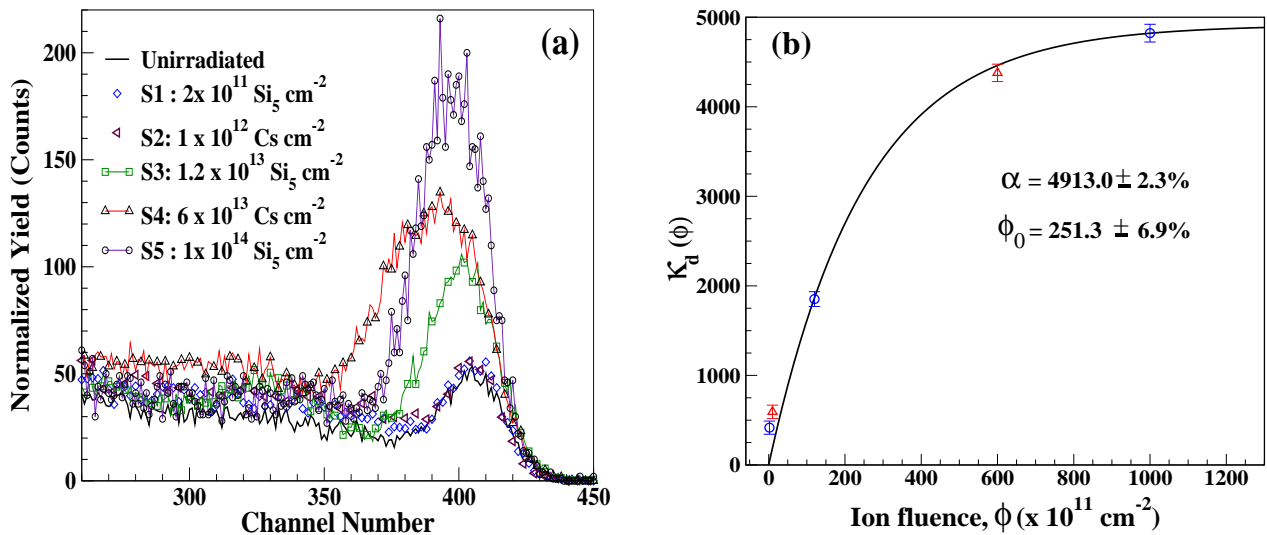

\begin{center}
\includegraphics[width=8.0cm,height=7.0cm]{fig1a_new.eps} %
\hspace{0.5cm}
\includegraphics[width=8.0cm,height=7.0cm]{fit-Chn-CsSi5.eps}
\label{fig1}
\caption{(Color online) (a) CRBS spectra for Si(100), 
implanted with 25 keV Si$_5^-$ and Cs$^-$ for different 
values of implantation fluence together with the same for 
a virgin sample. (b) Fluence ($\phi$) dependence of 
displaced Si atoms, $\kappa_d(\phi)$, for various 
implantations. Circles and triangles represent data for 
Si$_5$ and Cs implantations respectively. The continuous 
curve is a fit to the data points with clusters.}
\end{center}
\end{figure*} 

Cleaned Si(100) wafers ( $p$-type, 1-2.5 $\Omega cm$ ) were 
irradiated with 25 keV singly charged negative ions {\it viz} 
Si$_5^-$ and Cs$^-$ from a SNICS-II ion source (NEC, USA) 
using a low energy ion implanter facility. Mass analysis of 
the cluster ions was carried out using a $45^\circ$ sector 
magnet (ME/q$^2$ = 18 MeV amu). The base pressure in the 
target chamber during irradiations was maintained around 
2$\times 10^{-7}$ mbar. All the irradiations were carried 
out at room temperature with a beam flux of 2-3$\times 10^{10}$ 
ions cm$^{-2}$sec$^{-1}$ (ion current of $2-3 nA$) at 
$\sim$7$^\circ$ off the sample normal. In each case one 
part of the sample was kept unimplanted to serve as a 
reference. Five samples named S1-S5 were systematically 
irradiated with ions of similar mass (Si$_5^-$ or Cs$^-$) 
with gradually increasing ion fluence from 2$\times 10^{11}$ 
cm$^{-2}$ to 1$\times 10^{14}$ cm$^{-2}$. Three of these, 
{\it viz} S1, S3 and S5 were irradiated using Si$_5^-$  
clusters to fluences of $2\times10^{11}$ cm$^{-2}$, 
$1.2\times10^{13}$ cm$^{-2}$ and $1\times10^{14}$ cm$^{-2}$ 
respectively. The remaining two samples, S2 and S4 were 
irradiated with 25 keV Cs$^-$ ions to fluences of 
$1\times10^{12}$ cm$^{-2}$ and $6\times10^{13}$ 
cm$^{-2}$ respectively. These data are shown in Table. 1.

\begin{table}[h]
\caption{Sample names, ions used and integrated fluence.}
\begin{tabular}{ccc}
\hline
\hline
Samples&fluence &Ion species of \\
&&25 keV total energy\\
\hline
S1  & $2\times10^{11}$ cm$^{-2}$  & Si$_5^-$ \\
S2  & $1\times10^{12}$ cm$^{-2}$  & Cs$^-$     \\
S3  & $1.2\times10^{13}$cm$^{-2}$ & Si$_5^-$ \\
S4  & $6\times10^{13}$ cm$^{-2}$  & Cs$^-$     \\
S5  & $1\times10^{14}$ cm$^{-2}$  & Si$_5^-$ \\
\hline
\hline 
\end{tabular}
\end{table}

CRBS measurements were carried out on all the samples 
with 1.35 MeV He$^+$ with a Si surface barrier detector 
placed at 130$^\circ$ relative to the incident beam 
direction. The measurements were carried out at a 
steady beam current of 5 $nA$, using the 3 MV Pelletron 
accelerator (9SDH2, NEC, USA) facility at IOP, Bhubaneswar. 
In case of unirradiated Si(100), the reduction in the 
integrated total yield from random to a channeled spectrum 
was found to be $\sim$5$\%$. 

Following irradiation, the surface topography was examined 
by AFM in the tapping mode, using a multi-mode scanning probe 
microscope (Nanoscope IV, Veeco, USA). Measurements were 
performed in ambient condition using a Si cantilever with a 
nominal tip radius less than $\sim10~nm$. Image processing 
and analysis of the AFM data were carried out using the 
standard WSxM software package \cite{wsxm,wsxm-paper}.

\begin{figure*}
\begin{center}
\includegraphics[width=7.0cm,height=7.0cm]{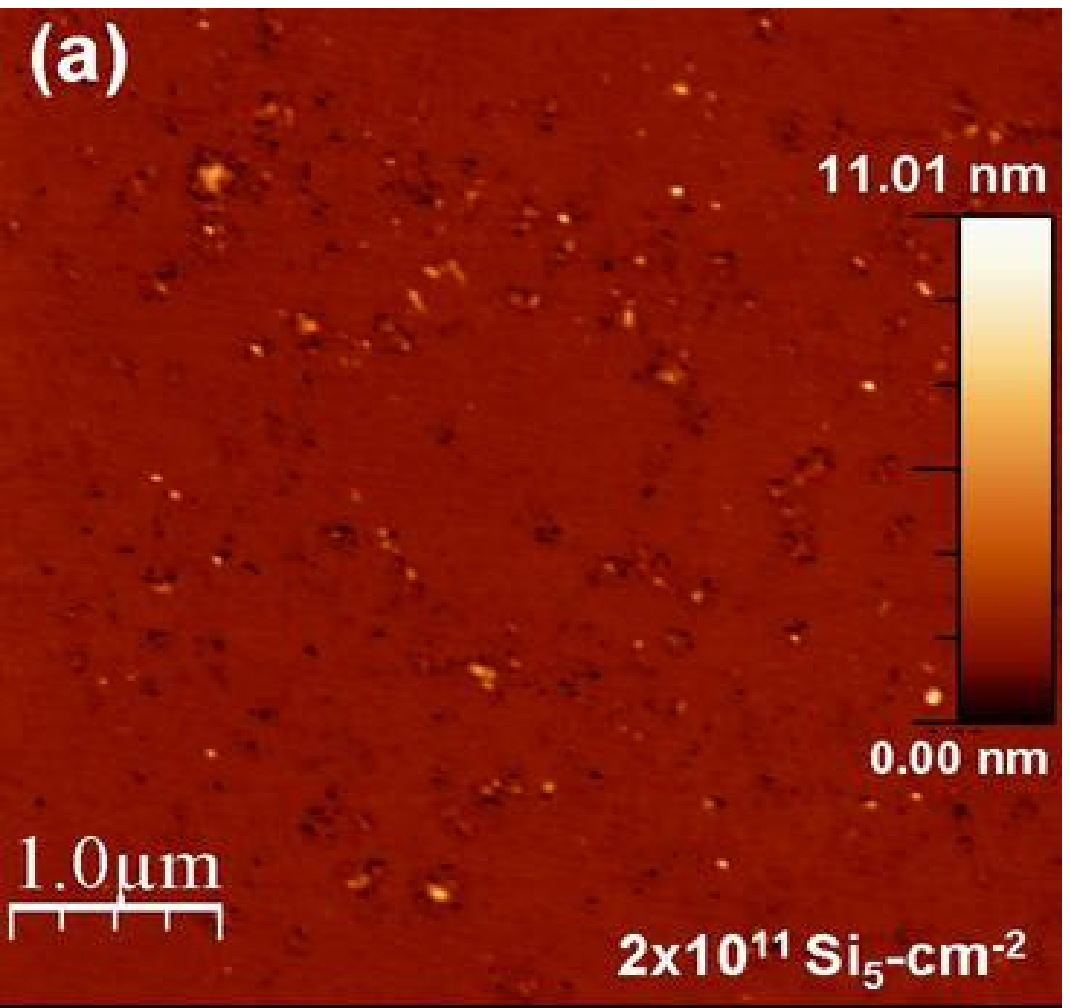} %
\hspace{0.5cm}
\includegraphics[width=7.0cm,height=7.0cm]{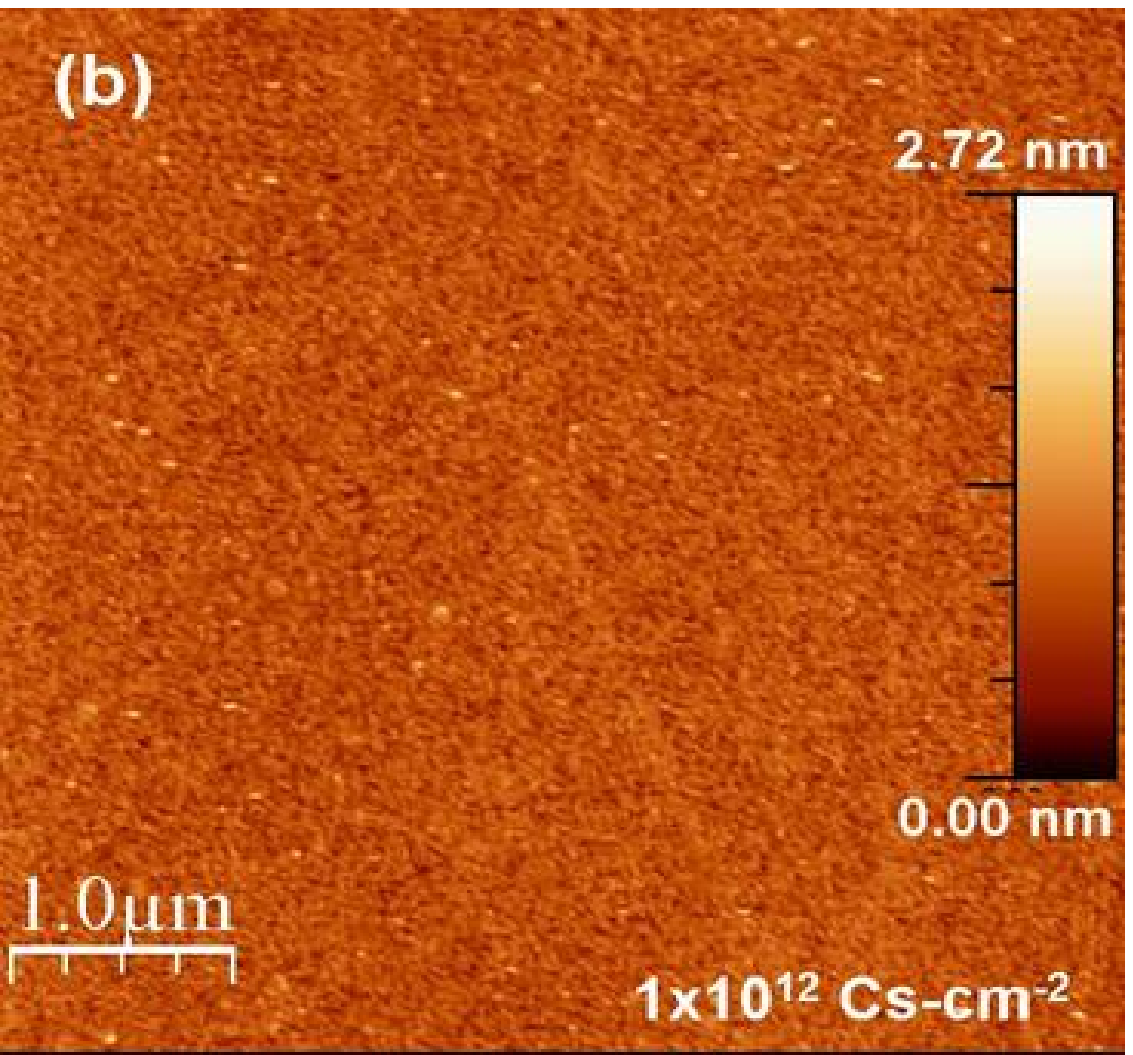}
\label{fig2}
\caption{(Color online) Surface morphology of samples 
(a) S1 and (b) S2. Both samples have the same monomer fluence 
as that of S2, {\it i.e.} $1\times 10^{12}$ atoms cm$^{-2}$.}
\end{center}
\end{figure*}
      
\section{Results and Discussion}

The CRBS results as measured for all the five samples 
{\it viz.} S1-S5 and a virgin sample (unirradiated area), 
are presented in Fig. 1(a). From the figure, one can 
observe a gradual increase in surface peak intensity 
with increase in ion fluence, $\phi$. This increase in 
the area of the surface peak, over and above that in the 
virgin sample, indicates growth in damage produced
with increase in the number of displacements. In line 
with this, the lowest fluence irradiation is found to 
induce very little damage, too small to be seen through 
CRBS data. However, increase in irradiation fluence from 
S1 to S3 (with cluster ions) and that for S2 to S4 (for 
Cs ions) are seen to result in enhancements of the surface 
peak, indicating an increase in defect production near 
the surface. For the sample S5, with the highest cluster 
irradiation fluence, the surface peak is seen to be the 
most intense. This sample, with a total single atom 
fluence of $5\times 10^{14}$ cm$^{-2}$ is expected 
to have an amorphous layer with a thickness 
of $\sim 8 nm$ extending from the surface \cite{agarwal}.
Cs$^-$ irradiation to a fluence, 0.6 times as that of
S5, is seen to produced a damage distribution going 
deeper into the bulk. 

Now we look at the difference in the surface peak areas 
between the irradiated and the unirradiated regions, 
which at lower irradiation fluence, is proportional to 
the number of displaced Si atoms, in the irradiated 
lattice. 
We denote this as $\kappa_d$ which
has been estimated for various samples integrating data 
(fig 1 (a)) between channels 340 and 440. Fig. 1(b) shows 
ion fluence, $\phi$, dependence of $\kappa_d$ for various 
implantation fluence. From the figure, one can see a clear 
nonlinear growth and saturation in damage with increase 
in ion fluence. Similar results, regarding a nonlinear 
growth in damage production and saturation has been seen 
earlier in case of 5 keV/atom C$_n$ cluster irradiations of 
Si(100) \cite{Cn-nimb07}. 

To get an idea about damage saturation and the amorphisation 
threshold, as has been done earlier \cite{Cn-nimb07,dobeli}, 
we have fitted the three points as obtained for Si$_5$ 
irradiation to an equation of the form, 

\begin{equation} 
\kappa_d(\phi)=\alpha (1-e^{-{\phi}/{\phi_0}})
\end{equation}

\noindent{}where $\alpha$ is a constant and $\phi_0$ 
corresponds to the threshold fluence for damage saturation. 
The best fit, as indicated by the smooth curve in Fig. 1(b), 
yields a value of $\phi_0$ equal to $(2.5\pm 7\%)\times 10^{13}$ 
cm$^{-2}$. One can see, at a fluence of $6\times 10^{13}$ 
cm$^{-2}$, the $\kappa_d$ value corresponding to 25 keV 
Cs atoms is almost the same as that expected for a similar 
mass Si$_5$ cluster at the same energy. This means, the heavy 
single atom induced cascades produce almost the same 
amount of damage or defects as 
those generated by a similar mass cluster ion. 
At this high fluence, because of overlapping of damage 
produced cluster induced nonlinear effects are difficult 
to be detected. 
However, in this case there are defects extending into the 
bulk while those for a cluster ion are better confined in a 
surface layer.

\begin{figure*} 
\begin{center}
\includegraphics[width=7.0cm,height=7.0cm]{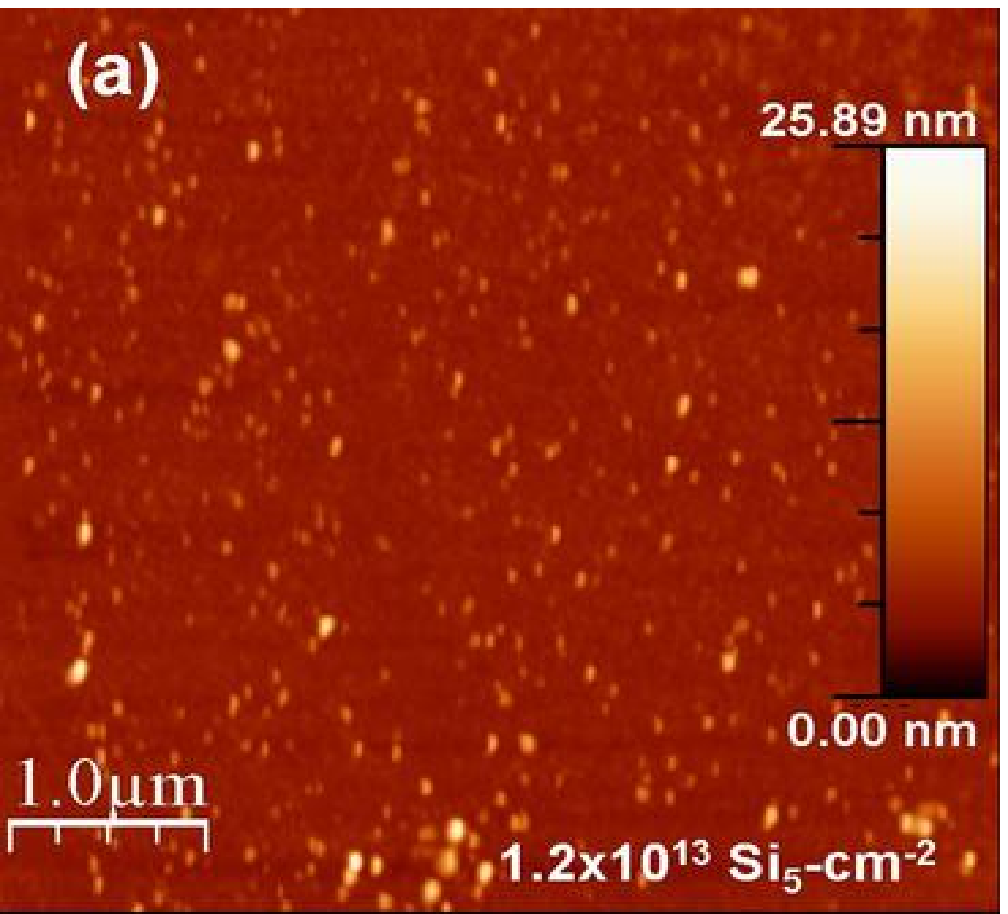} %
\hspace{0.5cm}
\includegraphics[width=7.0cm,height=7.0cm]{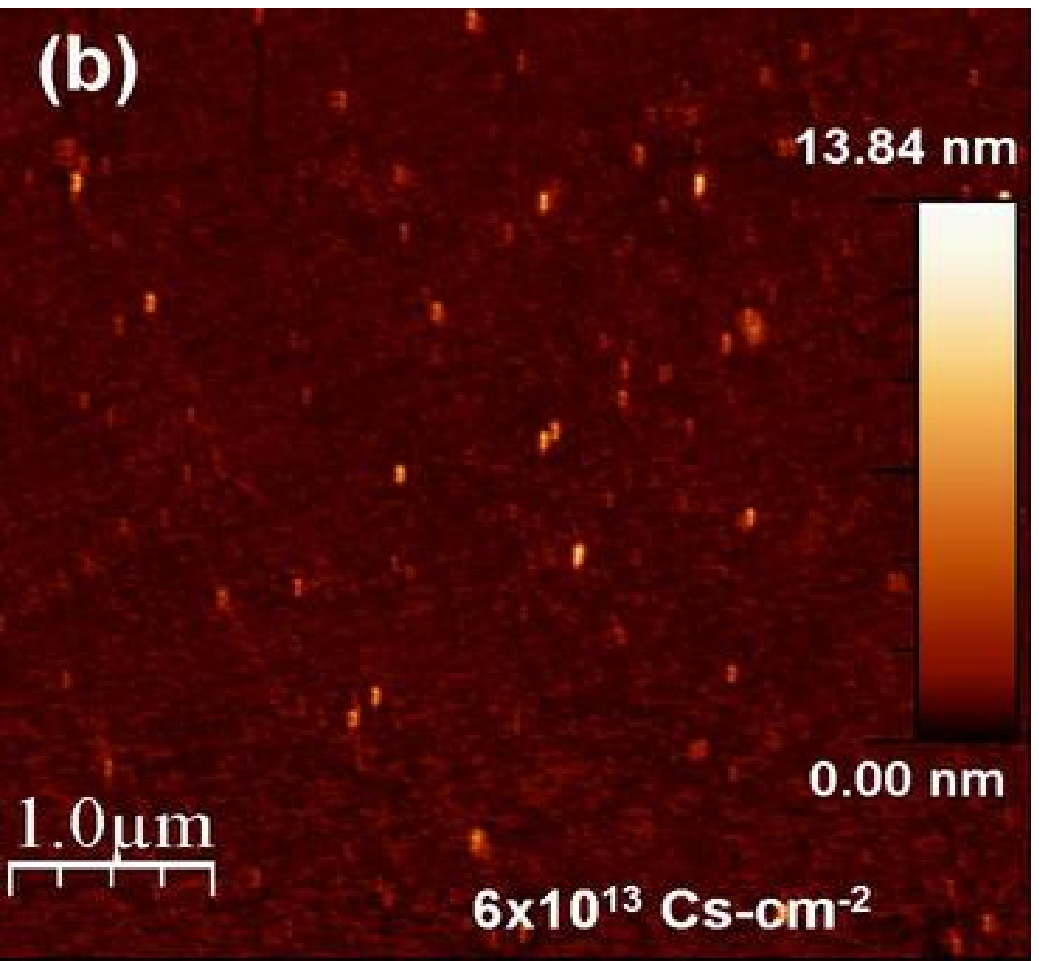}
\label{fig3}
\caption{(Color online) Surface morphology of samples 
(a) S3 and (b) S4. Both samples have the same monomer fluence 
as that of S3, {\it i.e.} $6\times 10^{13}$ atoms cm$^{-2}$.}
\end{center}
\end{figure*}

The above $\phi_0$ value of $2.5\times 10^{13}$ cm$^{-2}$ 
as obtained for $Si_5$ clusters, where damage saturation 
starts, corresponds to a total atomic fluence of 
$\sim 1.25\times 10^{14}$ cm$^{-2}$. This agrees 
with the finding of Agarwal {\it et al.} \cite{agarwal} who 
have shown the amorphisation threshold, for 5 keV Si in Si, 
to be $1-3\times 10^{14}$ cm$^{-2}$. In fact, with clusters, 
as expected the present value agrees with the lower limit. 
In view of this, at an Si$_5$ cluster fluence of 
$1\times 10^{14}$ cm$^{-2}$ (corresponding to an atomic 
fluence of $5\times 10^{14}$ cm$^{-2}$), we are already well 
above the threshold for amorphisation. 

\subsection{Surface features and AFM data}

As has been mentioned earlier, AFM has been used to 
study the surface topography in various samples. Some 
AFM pictures (top view), for ($5~\mu m\times 5~\mu m$) 
scanned areas, taken on the samples are shown in Figs. 
2 and 3. Figs 2(a) and (b) correspond to samples S1 and 
S2, while Figs. 3(a) and (b) correspond to samples S3 and 
S4 respectively. 

Usually the features on irradiated surfaces are described 
through a height-height correlation function which contains 
three important roughness parameters: (i) the vertical 
correlation length $\sigma$, (ii) the lateral correlation 
length $\xi$ and (iii) the roughness exponent $\alpha$
\cite{eklund,bray,fenner,fang1}. The lateral correlation 
length, $\xi$, describes the lateral characteristics of 
the surface, the roughness exponent $\alpha$ describing 
the static scaling properties. The most commonly reported 
parameter of surface roughness i.e. $\sigma$, or the 
root-mean-square (rms) roughness, characterizes the surface, 
only along the vertical direction. This is defined as 
standard deviation of the surface height profile, $h(x,y)$, 
at each point ($x ,y$) of a reference surface plane from 
the mean height ($<h>$), as given by,  

\begin{equation}
\sigma = \left[{1\over N}\sum_{i~=~1}^{N} {(h_i - <h>)^2}\right]^{1/2} 
\end{equation}

\noindent{}where $N$ is the number of pixels, $h_i = h(x,y)$ 
being the height at the $i^{th}$ pixel. 

In case of the sample S1, as shown in Fig 2(a), one can 
clearly see black dots corresponding to nanometer sized 
pits and bright spots corresponding to hillocks (color 
bars indicating the heights). The observed hillocks are 
seen to have heights ranging between $1-6~nm$ with average 
height of $1.64 nm$. It has an rms roughness, $\sigma$ of 
$0.34~nm$. Compared to this, there was hardly any surface 
feature in the case of a pristine Si(100) sample (AFM 
data not shown here). The rms roughness, $\sigma$, of the 
pristine sample, was found to be $0.13~nm$, which is about 
one third of that for S1, corresponding to a smooth polished 
surface. 
\begin{figure*}
\begin{center}
\includegraphics[width=8.0cm,height=7.5cm]{fig4a_final.eps} %
\hspace{0.5cm}
\includegraphics[width=8.0cm,height=7.5cm]{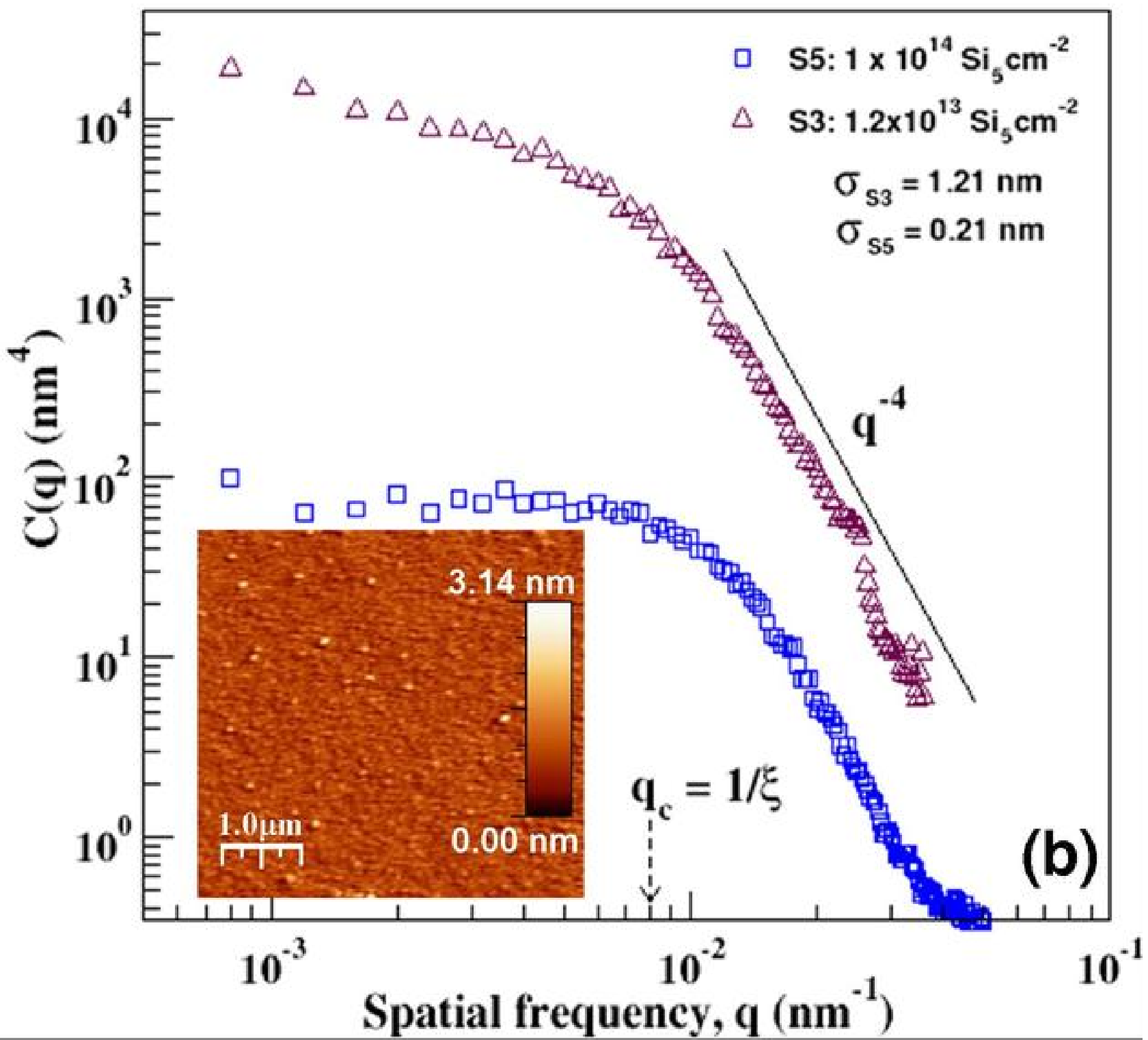}
\label{fig4}
\caption{(Color online) (a) $C(q)$, as a function of spatial 
frequency $q$, for Si$_5$ and Cs irradiated Si(100) at different 
irradiation fluence. $\bigcirc$, un-irradiated Si; 
$\bigtriangledown$, S1 (with Si$_5$ at $2\times10^{11} cm^{-2}$); 
$\Box$, S2 (with Cs monomer fluence $1\times10^{12} cm^{-2}$); 
$\bigtriangleup$ S3 (with Si$_5$ at $1.2\times10^{13} cm^{-2}$); 
$\diamondsuit$, S4 (with Cs at $6\times10^{13}$ cm$^{-2}$). 
(b) $C(q)$ for samples S3 and S5. Inset shows the AFM image for S5 for
a scan size of $5\mu m\times 5\mu m$.}
\end{center}
\end{figure*} 

In case of the lowest fluence Cs implanted sample, S2, 
the $\sigma$ for surface roughness was found to be 
$0.21~nm$, lying between that for S1 ($0.34 nm$) and 
the pristine sample ($0.13 nm$). What is important to 
see here is that Cs implantation to a five times higher 
fluence is not able to generate a surface roughness as 
observed in case of S1. Compared to the above, S3, 
implanted with Si$_5$ clusters, to a fluence sixty times that of the
sample S1, was found to have an rms roughness, $\sigma$, of
$1.21~nm$. However, the above sixty-fold increase in Si$_5$ cluster
fluence from S1 to S3, did not result in a proportionate increase in
the number of nanohillocks in S3. The surface topography of the sample
S4, implanted with Cs to a fluence five times that of S3 is shown in
Fig. 3(b). However, a much smaller number of nanohillocks, compared to
the S3 sample, could be seen. It has a $\sigma$ value of $0.42 nm$
which is significantly smaller than that observed for S3. The enhanced
surface roughness as seen with Si$_5$ implantations in S1 as compared
to S2 (and S3 as compared to S4), even with lower fluence of
irradiations, are primarily due to molecular effects coming from
overlapping of collision cascades of constituent atoms. Such effects
are absent during a single mass Cs irradiation which results in lower
damage accumulation and lower surface roughness. Compared to all the
above, the $\sigma$ value of surface roughness observed for the sample
S5, which had the highest cluster ion fluence, turned out to be $0.21
nm$. This was quite small compared to the same obtained for S3,  
irradiated with an almost one order of magnitude lower 
fluence. 

However, the parameter $\sigma$ is not enough to give 
a full characterization of the surface because it is 
limited by its sensitivity only in the vertical direction.
For example, two images with exactly the same rms 
roughness values can have different surface morphologies 
\cite{fang2}. In view of this, a power spectral density 
(PSD) analysis is often used to look at surface features and 
their possible origin 
\cite{petri-dumas,Tong,duparre,senthilkumar}. 
The PSD analysis is accomplished by radially averaging the 
square magnitude of the coefficients of the two-dimensional 
Fourier Transform of the digitized surface profile $h(x,y)$ 
defined by 

\begin{equation}
C(q)= {{1\over L^2}\left|\int\int {d^2r\over
    2\pi}\exp^{-iq.r}<h(r)>\right|^2} \\  
\end{equation}

\noindent{}where $h(r) = h(x,y)$. Here $q$ is the spatial 
frequency in reciprocal space, $L^2$ is the scanned area 
of length $L$ and $h(r)$ is the height at the position $r$. 
The PSD, which is the Fourier transform of the 
height-height correlation function, thus turns out to be 
a function of only one parameter, spatial frequency, $q$.
Identification of various processes for surface transport 
\cite{herring} is carried out using the stochastic rate 
equations for the surface evolution \cite{Tong-SScL}. In this 
analysis \cite{mayr-averback}, 
$C(q)=2D(q)/[\Sigma {a_\gamma} q^{\gamma}]$,
where $D(q)$ is a term coming from noise correlation, 
$a_\gamma$ being expansion coefficients. The index $\gamma$ 
has values of 1,2,3 and 4 representing four modes of surface 
transport {\it viz.} {\it viscous flow}, {\it evaporation-condensation}, 
{\it volume diffusion} and {\it surface diffusion} respectively. This 
index is further related to the roughness scaling exponent, 
$\alpha$ through the relation $\gamma=2(\alpha+1)$.  

The PSD spectra, $C(q)$, for all the samples S1 to S4 and 
pristine, are shown in Fig. 4(a). One can see, there is an
increase in spectral strength in the order 
pristine, S2, S1, S4 and S3 where the pristine has the 
lowest and S3 has the highest values. This order is exactly 
the same if one orders the samples according to the $\sigma$ 
values for the surface roughness. S2 and S4 have five times 
higher ion fluence as compared to S1 and S3 respectively. 
But the the surface features produced are seen to be much 
reduced. 

Having seen the above general features we now try to look 
at other details of $C(q)$. One can see that the pristine 
sample ($\sigma  = 0.13 nm$) has a cut off value, $q_c$, which 
is about 0.001 nm$^{-1}$. This corresponds to a large 
correlation length, $\xi$ of 1 $\mu m$. 
Further, it has a $\gamma$ close to 1. 
The Cs implanted sample, S2, has also a smooth surface 
($\sigma=0.21 nm$). The corresponding $C(q)$ indicates 
surface modulations over a similar length scale as the 
pristine sample. Compared to this, the Si$_5$ cluster 
implanted S1 sample, with one fifth of the fluence as in S2, shows
modulations with a higher value of $q_c$ of the order of $0.006
nm^{-1}$. This indicates a correlation length, $\xi$, of the order of  
$170 nm$. As shown earlier, it has a $\sigma$ value of $0.34 nm$. The
Fourier index $\gamma$ is seen to be about 2.5. This yields an
$\alpha$ value of 0.25, indicating the surface to be self affine with
anisotropic scaling along lateral and perpendicular directions.

The $C(q)$ for the higher fluence Cs irradiated sample, 
S4, shows a $q_c$ which is almost similar to that for 
the lower fluence Si$_5$ implanted sample, S1, indicating 
a similar correlation length, $\xi$, of $170 nm$. However, 
it shows a $\gamma$ value of around 2.2 resulting in an 
$\alpha$ value $\sim 0.1$. With a $\sigma$ value of 
$0.42 nm$, it has a higher roughness. This means, with a 
300 times higher fluence, Cs ions generate similar surface 
features as obtained for a low dose Si$_5$ implanted sample, 
S1. Compared to this, the Si$_5$ cluster implanted sample 
S3, 
with a fluence which is one fifth of that in S4, shows 
a $\gamma$ value close to 4, indicating an $\alpha$ value 
close to unity. This indicates the surface modulations 
to be self similar. But this sample has the highest 
surface roughness, $\sigma$ of $1.21 nm$ with an average 
height of $5 nm$. 

Now we look at what happens when cluster fluence is 
increased to a value well beyond the amorphisation 
threshold as in case of S5. An AFM image of the top 
view of S5, taken with a ($5\mu m\times 5\mu m$) scan 
size is shown in Fig. 4(b). The $C(q)$ spectrum for S5 
is also shown along with that for S3 in the same figure. 
One can clearly see that increasing the cluster ion 
fluence from 1.2$\times 10^{13}$ to 1$\times 10^{14}$ 
cm$^{-2}$, in going from S3 to S5, has resulted in no 
further change in the $\gamma$ value which has been 
found to saturate at 4. But the surface modulations 
changed to have a $\sigma$ value of $0.21 nm$ with a 
correlation length $\xi$ $\sim 125 nm$. With an $\alpha$ 
value close to 1, the surface features in S5 are self 
similar indicating isotropic scaling. However, as 
compared to the S3 case, $C(q)$ shows a significant 
reduction in the magnitude. At $q_c$, the ratio between 
the two is about 33. This can also be seen using the 
formula $C(q)=[\alpha\sigma^2\xi^2/\pi]$ at $q_c=1/\xi$. 
Since S3 and S5 have almost same $\alpha$ as well as 
$\xi$ values, the ratio of the $C(q)$s turn out to be 
nearly the same as the ratio of the $\sigma^2$ for the 
two cases. This is seen to be $(1.21/0.21)^2$ which is 
just about right. It is important to mention that the 
surface modulations in S5 show a mean height of $0.6 nm$. 
The small value of $C(q)$ for S5, is therefore seen to 
be coming from a correlation between small but nearly 
equal heights with a small $\sigma$ value. Compared to 
this $C(q)$ in S3 comes from a correlation between higher 
heights, with an average value of $5 nm$, again with a 
much higher value of $\sigma$.
 
Earlier, 5 keV Si impact on Si has been shown to result 
in creation of amorphous pockets coming form local melting 
and rapid quenching \cite{delaRubia}. The stress produced can 
result in formation of pits and bumps on the surface. This 
is in addition to roughening resulting from sputtering. 
This local melting and the associated movement of atoms, at 
lower fluence as in S1, may be responsible for a $\gamma$ 
value between 2 and 3. It also leads to a higher surface 
roughness as compared to a pristine sample. At lower 
fluence there are sparsely distributed amorphous pockets in the
matrix. Increase in the implantation fluence results in a fast 
growth in the number of amorphous patches resulting in
a growth in surface roughness, $\sigma$. This also results
in a growth in mean height, $<h>$, of surface structures 
produced, resulting in a growth of height-height 
correlation. Merging of amorphised regions at higher fluence, results 
in a building up of stress from a large number of 
{\it bond defects}. Finally there could be stress relaxation 
as the damaged lattice becomes unstable. This could result 
in a transition to a state with smaller surface features 
which is achieved by an effective movement of atoms in a 
lateral direction. This could be the reason behind getting 
a $\gamma$ value close to 4 as proposed for surface 
diffusion. This is probably how a smooth amorphous to 
crystalline interface can occur in Si under high fluence 
ion irradiation \cite{holland,bai}. This way crystalline 
to amorphous transition in Si, upon ion irradiation, is more 
like a phase transition induced by an accumulation of 
sufficient number of defects which was also suggested by 
several groups earlier \cite{swanson,holland,motooka}. 
In the present case, onset of this occurs at a cluster 
fluence of around 2.5$\times 10^{13}$ cm$^{-2}$. For a 
much higher cluster fluence (as in S5), a continuous 
amorphous layer parallel to the surface is produced, 
leading to much reduced values for $\sigma$ and $C(q)$. 
It is therefore not surprising that the $\xi$, $\alpha$ 
and $\gamma$ values as obtained for S5 are very similar 
to those obtained for a-Si films \cite{bray}. This also 
confirms that the top surface of the S5 sample is actually 
amorphised, in agreement with channeling data (Fig. 1(b)) 
where saturation in damage production has been obtained. 
It is also important to realize that $\kappa_d$ for S4 is 
much higher than that in case of S3 (Fig. 1(b)). In fact 
it is closer to that in S5. However, complete amorphisation 
does not occur here because the defects produced by Cs 
implantation are distributed over a greater depth 
resulting in comparatively less defect accumulation near 
the surface. 

\section{Conclusion}

To conclude, we have carried out a systematic study of 
25 keV Si$_5^-$ implantation induced damage and surface 
modifications in Si(100) where a nonlinear growth in 
subsurface damage, with fluence, is observed. The damage 
produced by similar mass Cs$^-$ ions, of the same energy, 
is seen to be distributed over a greater depth leading to 
much reduced surface features. With Si$_5$ clusters, the 
threshold fluence for amorphisation of Si surface is found 
to be $2.5\times 10^{13}$ clusters cm$^{-2}$, in agreement 
with earlier published data. Most importantly, at higher 
cluster fluence a transition to an amorphous state resulting 
in a much reduced surface roughness is indicated.

\section{Acknowledgments}

The authors would like to thank A.K. Behera for the efficient 
running of the implanter facility and all the operators of 
accelerator laboratory, IOP for help during CRBS runs. The 
technical help regarding some preliminary AFM measurements and
checks by S. Varma of IOP and S. Bhattacharjee of IMMT,
Bhubaneswar are also gratefully acknowledged.

\end{document}